\input{aipcheck}

\edef\optionlist{%
   \variorefoptionifavailable        % this is either varioref or
   draft,%
   \psnfssproblemoption              % this is empty unless problems
   tnotealph}
\documentclass[\optionlist]{aipproc}

\newcommand{\MS}{\overline{\rm MS}}
\newcommand{\RS}{\rm RS}

\newcommand{\OS}{\rm OS}

\newcommand{\nn}{\nonumber}
\newcommand{\be}{\begin{equation}}
\newcommand{\ee}{\end{equation}}
\newcommand{\bea}{\begin{eqnarray}}
\newcommand{\eea}{\end{eqnarray}}

\def\als{\alpha_{s}}

\def\lQ{\Lambda_{\rm QCD}}

\def\simg{{\ \lower-1.2pt\vbox{\hbox{\rlap{$>$}\lower6pt\vbox{\hbox{$\sim$}}}}\ }}
\def\siml{{\ \lower-1.2pt\vbox{\hbox{\rlap{$<$}\lower6pt\vbox{\hbox{$\sim$}}}}\ }}

\layoutstyle{6x9}

\usepackage{ttct0001}

\usepackage{shortvrb}
\MakeShortVerb\|

\makeatother

\begin{document}

\author{Antonio Pineda}{
  address={Institut f\"ur Theoretische Teilchenphysik, 
        Universit\"at Karlsruhe, 
        D-76128 Karlsruhe, Germany },
  email={pineda@particle.uni-karlsruhe.de},
}

\title{Large order behavior in perturbation theory of the pole mass and the singlet static potential}

\date{}

\begin{abstract}We discuss upon recent progress in our knowledge of
        the large order behavior in perturbation theory of the pole
        mass and the singlet static potential. We also discuss about
        the renormalon subtracted scheme, a matching scheme between
        QCD and any effective field theory with heavy quarks where,
        besides the usual perturbative matching, the first renormalon
        in the Borel plane of the pole mass is subtracted.
\end{abstract}

\maketitle

\section{Mass normalization constant}
\label{secmas}
In this paper, we review some results obtained in Ref. \cite{polemass}. 

The on-shell (OS) or pole mass can be related to the $\MS$ renormalized mass by the series
\be
\label{series}
m_{\OS} = m_{\MS} + \sum_{n=0}^\infty r_n \als^{n+1}\,, 
\ee
where the normalization point $\nu=m_{\MS}$ is understood for $m_{\MS}$
 and 
the first three coefficients $r_0$, $r_1$ and $r_2$ are known 
\cite{GRA90} ($\als=\als^{(n_l)}(\nu)$, where $n_l$ is the number of
light fermions). The pole mass is also known to be IR
finite and scheme-independent at any finite order in $\als$ \cite{irfinite}. We then define the Borel transform 
\be\label{borel}
m_{\OS} = m_{\MS} + \int\limits_0^\infty\mbox{d} t \,e^{-t/\als}
\,B[m_{\OS}](t)
\,,
\qquad B[m_{\OS}](t)\equiv \sum_{n=0}^\infty 
r_n \frac{t^n}{n!} . 
\ee
The behavior of the perturbative expansion of Eq. (\ref{series}) at large
orders is dictated by the closest singularity to the origin of its
Borel transform, which happens to be located at
$t=2\pi/\beta_0$, where we define 
$$
\nu {d \als \over d
\nu}=-2\als\left\{\beta_0{\als \over 4 \pi}+\beta_1\left({\als \over 4
\pi}\right)^2 + \cdots\right\}
.$$  
Being more precise, the behavior of the Borel transform near the
closest singularity at the origin reads (we define $u={\beta_0 t \over 4 \pi}$)
\be
B[m_{\OS}](t(u))=N_m\nu {1 \over
(1-2u)^{1+b}}\left(1+c_1(1-2u)+c_2(1-2u)^2+\cdots \right)+({\rm
analytic\; term}),
\ee
where by {\it analytic term}, we mean a piece that we expect it to be
analytic up to the next renormalon ($u=1$). This dictates the behavior of the perturbative expansion at large orders to be 
\be\label{generalm}
r_n \stackrel{n\rightarrow\infty}{=} N_m\,\nu\,\left({\beta_0 \over 2\pi}\right)^n
\,{\Gamma(n+1+b) \over
\Gamma(1+b)}
\left(
1+\frac{b}{(n+b)}c_1+\frac{b(b-1)}{(n+b)(n+b-1)}c_2+ \cdots
\right).
\ee
The different 
 $b$, $c_1$, $c_2$, etc ... can
 be obtained from the procedure used in \cite{Benb} (see
 \cite{Benb,polemass} for the explicit expressions).  
We then use the idea of \cite{Lee1} and define the new function
\bea
D_m(u)&=&\sum_{n=0}^{\infty}D_m^{(n)} u^n=(1-2u)^{1+b}B[m_{\OS}](t(u))
\\
\nn
&
=&N_m\nu\left(1+c_1(1-2u)+c_2(1-2u)^2+\cdots
\right)+(1-2u)^{1+b}({\rm analytic\; term})
\,.
\eea
This function is singular but bounded at the first IR renormalon. Therefore, we
can expect to obtain an approximate determination of $N_m$ if we know the first
coefficients of the series in $u$ and by using 
\be
N_m\nu=D_m(u=1/2)
.
\ee
The first three coefficients: $D_m^{(0)}$, $D_m^{(1)}$ and $D_m^{(2)}$
are known in our case. In order the calculation to make sense, we
choose $\nu \sim m$. For the specific choice $\nu=m$, we obtain (up to $O(u^3)$ with $u=1/2$)
\bea
\label{nm}
N_m&=&0.424413+0.137858+0.0127029= 0.574974 \quad (n_f=3)
\\
\nn
&=&0.424413+0.127505+0.000360952= 0.552279 \quad (n_f=4)
\\
\nn
&=&0.424413+0.119930-0.0207998= 0.523543 \quad (n_f=5)
\,.
\eea
The convergence is surprisingly good. The scale
dependence is also quite mild (see \cite{polemass}).

By using Eq. (\ref{generalm}), we can now go backwards and give some estimates for the $r_n$. They are displayed in Table \ref{tabm}. We can
see that they go closer to the exact values of $r_n$ when increasing
$n$. This makes us feel confident that we are near the asymptotic
regime dominated by the first IR renormalon and that for higher $n$ our
predictions will become an accurate estimate of the exact values. In
fact, they are quite compatible with the results obtained by other
methods like the large $\beta_0$ approximation (see Table \ref{tabm}).

We can now try to see how the large $\beta_0$ approximation works in
the determination of $N_m$. In order to do so, we study the one chain
approximation from which we obtain the value \cite{BenekeBraunren} 
\be
N_m^{({\rm large}\; \beta_0)}={C_f \over \pi}e^{5 \over 6}=0.976564.
\ee 
By comparing with Eq. (\ref{nm}), we can see that it does not
provide an accurate determination of $N_m$. This may seem to be in
contradiction with the accurate values that the large $\beta_0$
approximation provides for the $r_n$ (starting at $n=2$) in Table
\ref{tabm}.  Lacking of any physical explanation for this fact, it
may just be considered to be a numerical accident. In fact, the agreement
between our determination and the large $\beta_0$ results does not
hold at very high orders in the perturbative expansion, whereas we
believe, on physical grounds since our approach incorporates the
exact nature of the renormalon, that our determination should go
closer to the exact result at high orders in perturbation
theory. Nevertheless, the large $\beta_0$ approximation remains
accurate up to relative high orders.

\begin{table}
\begin{tabular}{lrrrrr}
\hline
   \tablehead{1}{l}{b}{${\tilde r}_n=r_n/m_{\MS}$}
  & \tablehead{1}{r}{b}{${\tilde r}_0$}
  & \tablehead{1}{r}{b}{${\tilde r}_1$}
  & \tablehead{1}{r}{b}{${\tilde r}_2$}
  & \tablehead{1}{r}{b}{${\tilde r}_3$}   
  & \tablehead{1}{r}{b}{${\tilde r}_4$}\\
\hline
exact ($n_f=3$) & 0.424413 & 1.04556 & 3.75086  & --- &
 ---  \\
Eq. (\ref{generalm}) ($n_f=3$) & 0.617148  & 0.977493 &  3.76832
  & 18.6697  &  118.441  \\
large $\beta_0$ ($n_f=3$) & 0.424413  & 1.42442 &  3.83641
  & 17.1286   &  97.5872   \\ \hline
exact ($n_f=4$) & 0.424413 & 0.940051 & 3.03854 & --- 
 & ---   \\
Eq. (\ref{generalm}) ($n_f=4$) & 0.645181 & 0.848362 & 3.03913
 & 13.8151 & 80.5776   \\ 
large $\beta_0$ ($n_f=4$) & 0.424413 & 1.31891 &  3.28911
  & 13.5972  & 71.7295    \\ \hline
exact ($n_f=5$) &0.424413 & 0.834538  & 2.36832 & --- & ---  \\
Eq. (\ref{generalm}) ($n_f=5$) & 0.706913 & 0.713994 & 2.36440 &
 9.73117  & 51.5952 \\
large $\beta_0$ ($n_f=5$) & 0.424413  & 1.21339 & 2.78390 
  & 10.5880  & 51.3865   \\  \hline
\end{tabular}
\caption{Values of $r_n$ for $\nu=m_{\MS}$. Either the exact result, the estimate using Eq. (\ref{generalm}), or the estimate using the large
  $\beta_0$ approximation \cite{BenekeBraun}.}
\label{tabm}
\end{table}

\section{Static singlet potential normalization constant}
\label{secpot}

One can think of playing the same game with the singlet 
static potential in the situation where $\lQ \ll 1/r$. Its perturbative expansion reads 
\be
V_s^{(0)}(r;\nu_{us})=\sum_{n=0}^\infty V_{s,n}^{(0)} \als^{n+1}.
\ee
The first
three coefficients $V_{s,0}^{(0)}$, $V_{s,1}^{(0)}$ and $V_{s,2}^{(0)}$ are known
\cite{FSP}. At higher orders in perturbation theory the log dependence on the IR cutoff $\nu_{us}$ appears \cite{short}. Nevertheless, these logs are not
associated to the first IR renormalon (see \cite{polemass}), so
we will not consider them further in this section. We now use the observation that the first IR renormalon of the singlet
 static potential cancels with the renormalon of (twice) the pole
 mass. We can then read the asymptotic behavior of the static potential from the one
 of the pole mass and work analogously to the
 previous section. We define the Borel transform 
\be\label{borelb}
V_s^{(0)} = \int\limits_0^\infty\mbox{d} t \,e^{-t/\als}\,B[V_s^{(0)}](t)
\,,
\qquad 
B[V_s^{(0)}](t)\equiv \sum_{n=0}^\infty 
V_{s,n}^{(0)} \frac{t^n}{n!} . 
\ee
The closest singularity to the origen is located at 
$t=2\pi/\beta_0$. This dictates the behavior of the perturbative expansion at large orders to be 
\be\label{generalV}
V_{s,n}^{(0)} \stackrel{n\rightarrow\infty}{=} N_V\,\nu\,\left({\beta_0 \over 2\pi}\right)^n
 \,{\Gamma(n+1+b) \over
 \Gamma(1+b)}
\left(
1+\frac{b}{(n+b)}c_1+\frac{b(b-1)}{(n+b)(n+b-1)}c_2+ \cdots
\right)
,
\ee
and the Borel transform near the singularity reads
\be
B[V_{s}^{(0)}](t(u))=N_V\nu {1 \over
(1-2u)^{1+b}}\left(1+c_1(1-2u)+c_2(1-2u)^2+\cdots \right)+({\rm
analytic\; term}).
\ee
In this case, by {\it analytic term}, we mean an analytic function up to
 the next IR renormalon at $u=3/2$. 

As in the previous section, we define the new function
\bea
D_V(u)&=&\sum_{n=0}^{\infty}D_V^{(n)} u^n = (1-2u)^{1+b}B[V_{s}^{(0)}](t(u))
\\
\nn
&
=&N_V\nu\left(1+c_1(1-2u)+c_2(1-2u)^2+\cdots
\right)+(1-2u)^{1+b}({\rm analytic\; term})
\eea
and try to obtain an approximate determination of $N_V$ by using the
 first three (known) coefficients of this series. By a discussion
 analogous to the one in the previous section, we fix $\nu=1/r$. We
 obtain (up to $O(u^3)$ with $u=1/2$)
\bea
\label{nv}
N_V&=&-1.33333+0.571943-0.345222 = -1.10661 \quad (n_f=3) \\ \nn
&=&-1.33333+0.585401-0.329356 = -1.07729 \quad (n_f=4) \\ \nn
&=&-1.33333+0.586817-0.295238 = -1.04175 \quad (n_f=5) \,.  
\eea 
The
convergence is not as good as in the previous section. Nevertheless,
it is quite acceptable and, in this case, apparently, we have a sign
alternating series. In fact, the scale dependence is quite mild (see \cite{polemass}). Overall,
up to small differences, the same picture than for $N_m$ applies.

So far we have not made use of the fact that $2N_m+N_V = 0$. We use
this equality as a check of the reliability of our calculation. We can see
that the cancellation is quite dramatic. We obtain
\begin{eqnarray*} 
2{2N_m+N_V \over 2N_m-N_V} = \,\left\{
\begin{array}{ll}
\displaystyle{ 0.038 }&\ \ , \, n_f=3 \\
\displaystyle{0.025}&\ \ ,\, n_f=4 \\
\displaystyle{0.005}&\ \ ,\, n_f=5.
\end{array} \right.
\end{eqnarray*}
We can now obtain estimates for $V_{s,n}^{(0)}$ by using Eq.
(\ref{generalV}).  They are displayed in Table \ref{tabv}. Note that in
Table \ref{tabv} no input from the static potential has been used since
even $N_V$ have been fixed by using the equality $2N_m=-N_V$. We can see
that the exact results are reproduced fairly well (the same discussion
than for the $r_n$ determination applies). This makes us feel confident
that we are near the asymptotic regime dominated by the first IR
renormalon and that for higher $n$ our predictions will become an
accurate estimate of the exact results. The
comparison with the values obtained with the large $\beta_0$
approximation would go (roughly) along the same lines than for the mass
case, although the large $\beta_0$ results seem to be less accurate in
this case (see Table \ref{tabv}).
\begin{table}
\begin{tabular}{lrrrrr}
\hline
   \tablehead{1}{l}{b}{${\tilde V}_{s,n}^{(0)}= r V_{s,n}^{(0)}$}
  & \tablehead{1}{r}{b}{${\tilde V}_{s,0}^{(0)}$}
  & \tablehead{1}{r}{b}{${\tilde V}_{s,1}^{(0)}$}
  & \tablehead{1}{r}{b}{${\tilde V}_{s,2}^{(0)}$}
  & \tablehead{1}{r}{b}{${\tilde V}_{s,3}^{(0)}$}   
  & \tablehead{1}{r}{b}{${\tilde V}_{s,4}^{(0)}$}\\
\hline
exact ($n_f=3$) & -1.33333 & -1.84512 & -7.28304  & --- &
 ---  \\
Eq. (\ref{generalV}) ($n_f=3$) & -1.23430 & -1.95499 & -7.53665
  & -37.3395  & -236.882  \\
large $\beta_0$ ($n_f=3$) & -1.33333  & -2.69395 & -7.69303
  & -34.0562   &  ---  \\
\hline
exact ($n_f=4$) & -1.33333 & -1.64557 & -5.94978 & --- 
 & ---   \\
Eq. (\ref{generalV}) ($n_f=4$)  & -1.29036 & -1.69672 & -6.07826
 & -27.6301  & -161.155   \\ 
large $\beta_0$ ($n_f=4$) & -1.33333 & -2.49440  &  -6.59553
  & -27.0349  & ---  \\ 
 \hline
exact ($n_f=5$) & -1.33333 & -1.44602 & -4.70095 & --- & ---  \\
Eq. (\ref{generalV}) ($n_f=5$) &-1.41383 & -1.42799 & -4.72881 &
  -19.4623 & -103.190  \\
large $\beta_0$ ($n_f=5$) & -1.33333  &-2.29485  &  -5.58246
  & -21.0518  & --- \\
 \hline
\end{tabular}
\caption{Values of $V_{s,n}^{(0)}$ with $\nu=1/r$. Either the exact result (when available), the
  estimate using Eq. (\ref{generalV}), or the estimate using the large
  $\beta_0$ approximation \cite{KSH}.}
\label{tabv}
\end{table}

In order to avoid large corrections from terms depending on $\nu_{us}$,
the predictions should be understood with $\nu_{us}=1/r$. 

\section{Renormalon subtracted scheme}
\label{secdefRS}
 
In effective theories with heavy quarks, the inverse of the heavy
quark mass becomes one of the expansion parameters (and matching
coefficients). A natural choice in the past (within the infinitely
many possible definitions of the mass) has been the pole mass because
it is the natural definition in OS processes where the particles
finally measured in the detectors correspond to the fields in
the Lagrangian (as in QED). Unfortunately, this is not the case in QCD
and one reflection of this fact is that the pole mass suffers from renormalon singularities. Moreover, these renormalon singularities lie
close together to the origin and perturbative calculations have gone
very far for systems with heavy quarks. At the practical level, this
has reflected in the worsening of the perturbative expansion in
processes where the pole mass was used as an expansion parameter. It
is then natural to try to define a new expansion parameter replacing
the pole mass but still being an adequate definition for threshold
problems. This idea is not new and has already been pursued in the
literature, where several definitions have arisen \cite{kinetic}. We can not resist the tentation of trying our own
definition. We believe that, having a different systematics than the
other definitions, it could further help to estimate the errors in the
more recent determinations of the $\MS$ quark mass. Our definition, as the
definitions above, try to cancel the bad perturbative behavior
associated to the renormalon. On the other hand, we would like to
understand this problem within an effective field theory
perspective. From this point of view what one is seeing is that the
coefficients multiplying the (small) expansion parameters in the
effective theory calculation are not of
natural size (of $O(1)$). The natural answer to this problem is that
we are not properly separating scales in our effective theory and some
effects from small scales are incorporated in the matching
coefficients. These small scales are dynamically generated in $n$-loop
calculations ($n$ being large) and are of $O(m\,e^{-n})$ (we are
having in mind a large $\beta_0$ evaluation) producing the bad
(renormalon associated) perturbative behavior. In order to overcome this
problem, we may think of doing the Borel transform. In that case, the
renormalon singularities correspond to the non-analytic terms in
$1-2u$. These terms also exist in the effective theory.  Therefore,
our procedure will be to subtract the pure renormalon contribution in
the new mass definition, which we will call renormalon subtracted
(RS) mass, $m_{\RS}$.  We define the Borel transform of $m_{\RS}$ as
follows 
\be 
B[m_{\RS}]\equiv B[m_{\OS}] -N_m\nu_f {1 \over
(1-2u)^{1+b}}\left(1+c_1(1-2u)+c_2(1-2u)^2+\cdots \right), 
\ee 
where
$\nu_f$ could be understood as a factorization scale between QCD and
NRQCD (or HQET) and, at this stage, should be smaller than $m$. The
expression for $m_{\RS}$ reads 
\be
\label{mrsvsmpole}
m_{\RS}(\nu_f)=m_{\OS}-\sum_{n=0}^\infty  N_m\,\nu_f\,\left({\beta_0 \over
2\pi}\right
)^n \als^{n+1}(\nu_f)\,\sum_{k=0}^\infty c_k{\Gamma(n+1+b-k) \over
\Gamma(1+b-k)}
\,,
\ee
where $c_0=1$. 
We expect that with this renormalon free definition the 
coefficients multiplying the expansion parameters in the effective
theory calculation will have a natural size and also the coefficients multiplying
the powers of $\als$ in the perturbative expansion relating $m_{\RS}$ with $m_{\MS}$. Therefore,
we do not loose accuracy if we first obtain $m_{\RS}$ and later on we
use the perturbative relation between $m_{\RS}$ and
$m_{\MS}$ in order to obtain the latter. Nevertheless, since we will work order by order in
$\als$ in the relation between $m_{\RS}$ and
$m_{\MS}$, it is important to expand everything in terms of
$\als$, in particular $\als(\nu_f)$, 
in order to achieve the renormalon cancellation order by order in
$\als$. Then, the
perturbative expansion in terms of the $\MS$ mass reads 
\be
m_{\RS}(\nu_f)=m_{\MS} + \sum_{n=0}^\infty r^{\RS}_n\als^{n+1}\,,
\ee
where $r^{\RS}_n=r^{\RS}_n(m_{\MS},\nu,\nu_f)$. These $r^{\RS}_n$ are
the ones expected to be of natural
size (or at least not to be artificially enlarged by the first IR renormalon).

In Ref. \cite{polemass}, we have applied this scheme to potential
NRQCD and HQET. For the former, by using the $\Upsilon(1S)$ mass, we have obtained
a determination of the $\MS$ bottom quark mass. For the latter, we
have obtained a value of the charm mass by using
the difference between the $D$ and $B$ meson mass. In both cases the
convergence is significantly improved if compared with the analogous
OS evaluations.

\end{document}